# Optical Anisotropy and Phase Matching in Non-Centrosymmetric Perovskite Oxides from DFT+$U$ and DFT+$U$+$V$ Functionals

Mohamed S.M.M. Ali* and Ismaila Dabo

*Department of Materials Science and Engineering, and Wilton E. Scott Institute for Energy Innovation, Carnegie Mellon University, Pittsburgh, PA 15213, USA*

Optical anisotropy underpins the operation and performance of a broad range of photonic and quantum technologies. In this work, we critically examine the accuracy of density-functional theory approximations with onsite and intersite Hubbard corrections (the DFT+$U$ and DFT+$U$+$V$ functionals) in predicting the anisotropic optical response of the non-centrosymmetric perovskite oxides, such as $BaTiO_3$, $LiNbO_3$, $KNbO_3$, and $PbTiO_3$. It is found that correcting self-interaction errors using DFT+$U$ alone does not capture the optoelectronic response of these materials, often leading to a suppression of their optical anisotropy. While intersite Hubbard interactions restore this anisotropy, the choice of the (inter)atomic orbital manifold that defines the Hubbard correction remains critical to its accuracy. The predictive performance of the resulting, systematically validated DFT+$U$+$V$ functional is achieved at a fraction of the computational cost of hybrid functionals and many-body perturbation theory calculations. As benchmarks, we investigate Zn- and (Bi,Mn)-substituted $BaTiO_3$ solid solutions; the latter exhibit polarization-dependent bandgap narrowing from mid-gap states, substantially enhancing the dichroic ratio and birefringence with promising implications for polarization-sensitive photodetectors and integrated photonics.

## Introduction

Non-centrosymmetric perovskite oxides, such as $BaTiO_3$, $LiNbO_3$, $KNbO_3$, and $PbTiO_3$, are attractive for photonic applications due to their broken inversion symmetry and strong orbital hybridization that give rise to distinctive optical properties. In the nonlinear dielectric regime, the response of the polarization $\boldsymbol{P}$ can be expanded in powers of the applied electric field $\boldsymbol{E}$ as $\varepsilon_0^{-1} P_i = \chi_{ij}^{(1)} E_j + \chi_{ijk}^{(2)} E_j E_k + \chi_{ijkl}^{(3)} E_j E_k E_l + \cdots$, where $\boldsymbol{\chi}^{(n)}$ denotes the $n$th-order electric susceptibility tensor, $\varepsilon_0$ is the vacuum permittivity, and the summations run over all repeated spatial coordinates. Under a time-dependent electric field, the second- and higher-order terms generate polarization components at frequencies different from that of the incident field, producing nonlinear optical phenomena, such as second-harmonic generation and the electro-optic effect. As evident from the expansion of the polarization, inversion symmetry breaking is required for a non-zero second-order susceptibility. Non-centrosymmetric crystals also commonly exhibit birefringence, which plays a central role in optical modulation, phase matching for frequency conversion, and the generation of entangled photon pairs for quantum communication (Fig. 1) [1–6].

Predicting the optical properties of these materials from first principles requires an accurate description of their electronic structure and dielectric response. Density-functional theory (DFT) approximations provide an efficient computational approach to describe the optical behavior of materials, yet they systematically underestimate the bandgaps of most semiconductors and transition-metal oxides due to self-interaction errors inherent in semilocal exchange–correlation functionals [7–13]. Hybrid functionals and many-body perturbation theory methods substantially improve the description of electronic excitations and optical spectra; however, their computational cost may limit their applicability to materials screening and optimization [14–19].

Self-consistent Hubbard corrections provide a computationally efficient alternative to these approaches by mitigating self-interaction errors while retaining the scalability of DFT methods. In particular, the DFT+$U$+$V$ formalism incorporates both onsite and intersite Hubbard interactions, enabling a more realistic description of the covalent bonding that governs the optical response of transition-metal oxides [20–26]. Although recent studies have demonstrated the ability of self-consistent Hubbard functional, including first-principles Hubbard functionals parameterized using density-functional perturbation theory (DFPT) [21, 25] and the pseudohybrid ACBN0 functional [27–30], to predict structural properties and bandgaps, their accuracy in describing optical anisotropy and nonlinear optics remains largely unexplored.

In this work, we employ Hubbard corrections to density-functional functionals for simulating the optical anisotropy of ferroelectric oxides $BaTiO_3$, $LiNbO_3$, $KNbO_3$, and $PbTiO_3$ as benchmark materials. Since the accurate prediction of optical anisotropy requires the simultaneous description of both the electronic structure and the crystal lattice geometry, we parameterize Hubbard functionals non-empirically using DFPT and we discuss the accuracy of the calculated optical response according to the choice of the orbital manifolds to which the Hubbard corrections are applied. In addition, we compare the optical anisotropy obtained using DFPT and ACBN0. We then assess optical phase matching (the condition of minimal energy loss) in the materials considered and introduce sensitive metrics of phase-matching tunability. Finally, we extend our simulations to solid solutions and study the effect of doping on optical anisotropy using $BaTiO_3$ as a prototype.

*Corresponding author: mali3@andrew.cmu.edu

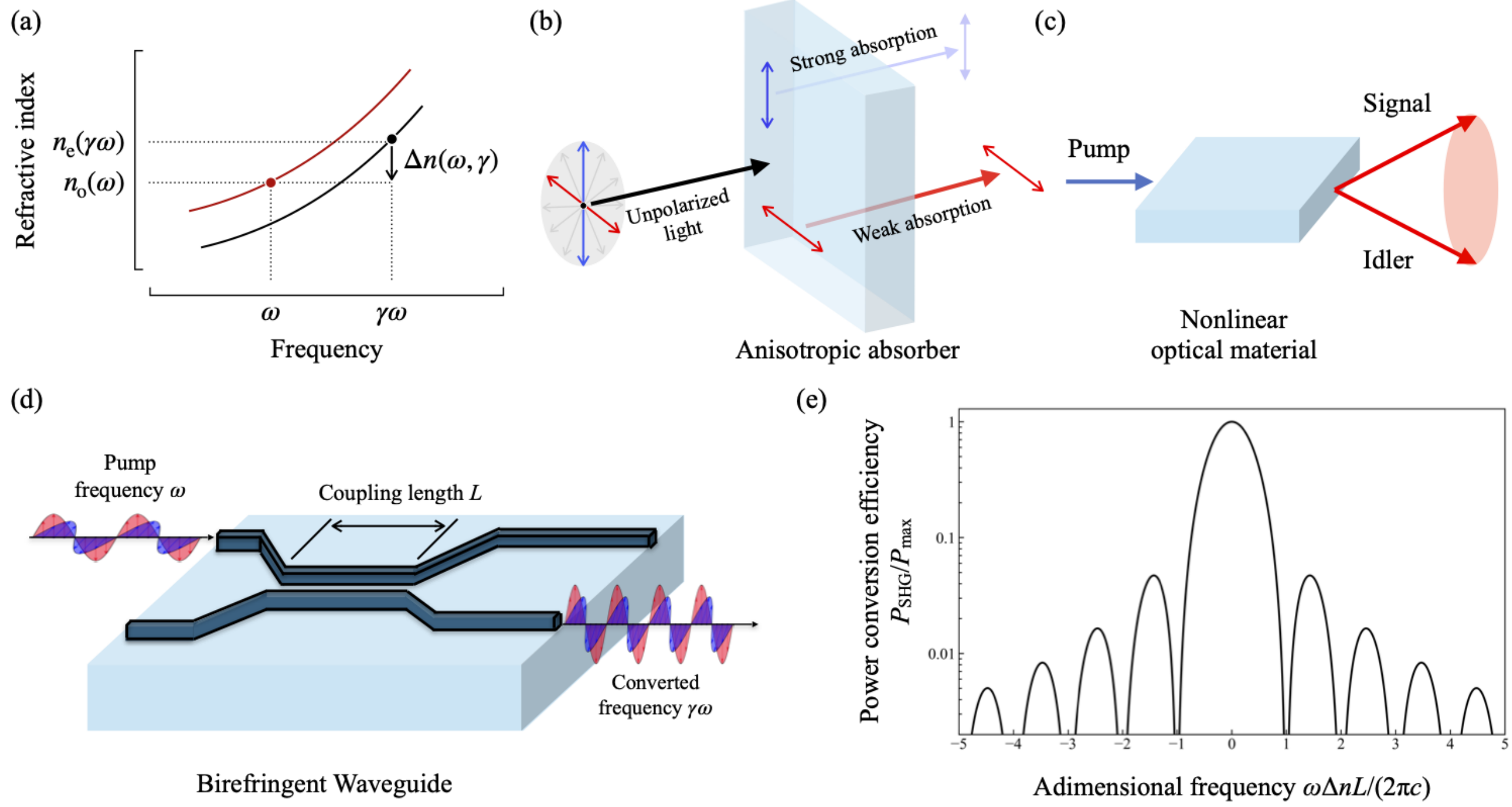

Figure 1: Optical anisotropy enables a broad range of applications. (a) Dispersion of a uniaxial crystal where $n_\mathrm{o}$ and $n_\mathrm{e}$ are the ordinary and extraordinary refractive indices, respectively. $\Delta n(\omega,\gamma) \equiv n_\mathrm{o}(\omega) - n_\mathrm{e}(\gamma\omega)$ denotes the frequency-scaled refractive-index mismatch, where $\gamma$ is the scaling factor ($\gamma = 2$ for SHG). (b) Materials with anisotropic absorption used for polarization-sensitive photodetectors. (c) Generation of polarization-entangled photon pairs for quantum communications by spontaneous parametric down-conversion, whereby a high energy pump photon splits into two lower energy photons (signal and idler). Birefringent phase matching, enabled by optical anisotropy, is one of the widely-used approaches for efficient photon-pair generation. (d) Birefringent waveguide platform for frequency conversion, whereby optical anisotropy compensates chromatic dispersion to enable efficient energy transfer between optical frequencies. (e) The SHG conversion efficiency in a birefringent phase-matched waveguide depends on the refractive-index mismatch $\Delta n(\omega,\gamma = 2)$, with $L$ representing the waveguide coupling length, $c$ the speed of light, and $P_\mathrm{SHG}/P_\mathrm{max}$ the normalized output power.

## Extended Hubbard Formalism

In the extended Hubbard (DFT+$U$+$V$) framework, a corrective term is added to the total density-functional energy of the system

$$E_{\mathrm{DFT}+U+V} = E_\mathrm{DFT} + E_{U+V}\,, \tag{1}$$

where the correction $E_{U+V}$ involves onsite $U^I$ and intersite $V^{IJ}$ Hubbard parameters that capture electrostatic and exchange–correlation interactions between electron pairs. To calculate the parameters $U^I$ and intersite $V^{IJ}$, we first define the intersite occupancy matrix between site $I$ and site $J$:

$$n^{IJ\sigma}_{mm'} = \sum_{nk} f_{nk\sigma} \langle \psi_{nk\sigma}|\phi^J_{m'}\rangle\langle\phi^I_m|\psi_{nk\sigma}\rangle, \tag{2}$$

where $\psi_{nk\sigma}$ are Kohn–Sham (KS) states with occupancies $f_{nk\sigma}$ projected onto a set of localized (atom-centered) orbitals $\phi^I_m$, $\sigma$ is the electron spin, $m$ and $m'$ are the angular-momentum quantum numbers, and $n^{IJ\sigma}$ is the matrix element between orbital $m$ at site $I$ and orbital $m'$ at site $J$.

We then write the general expression of the intersite interaction operator, which describes the energy cost of two electrons occupying orbitals at different sites $I$ and $J$ as

$$\hat{V}_\mathrm{int} = \frac{1}{2}\sum_{I\neq J}\sum_{\sigma\sigma'}\sum_{mm'} V^{IJ}\hat{c}^\dagger_{Im\sigma}\hat{c}^\dagger_{Jm'\sigma'}\hat{c}_{Jm'\sigma'}\hat{c}_{Im\sigma}, \tag{3}$$

where $\hat{c}^\dagger$ and $\hat{c}$ are the fermionic creation and annihilation operators, respectively. Within the KS framework, the interacting many-electron problem is mapped onto an auxiliary non-interacting system described by a Slater determinant. Therefore, Wick's theorem [31] applies exactly, allowing higher-order expectation values to be factorized into products of two-point correlation functions. This factorization yields an energy expression that depends only on the occupation matrices of the system:

$$E = \frac{1}{2}\sum_{I\neq J}\sum_{mm'\sigma\sigma'} V^{IJ}\Big[\langle \hat{c}^\dagger_{Im\sigma}\hat{c}_{Im\sigma}\rangle\langle \hat{c}^\dagger_{Jm'\sigma'}\hat{c}_{Jm'\sigma'}\rangle - \langle \hat{c}^\dagger_{Im\sigma}\hat{c}_{Jm'\sigma'}\rangle\langle \hat{c}^\dagger_{Jm'\sigma'}\hat{c}_{Im\sigma}\rangle\Big]. \quad (4)$$

Distributing the sum in the first term and collapsing the sum over $\sigma'$ in the second, we obtain

$$E = \frac{1}{2}\sum_{I\neq J} V^{IJ}\left[\left(\sum_{m\sigma} n^{I\sigma}_{mm}\right)\left(\sum_{m'\sigma'} n^{J\sigma'}_{m'm'}\right) - \sum_{\sigma}\sum_{mm'} n^{IJ\sigma}_{mm'} n^{JI\sigma}_{m'm}\right], \quad (5)$$

where the first term is the Hartree-like electrostatic interaction between the total charges at sites $I$ and $J$, and the second term is the Fock-like exchange between electrons with the same spin. Following Ref. [32], in the zero-configurational-width approximation, the effective counting energy $\tilde{E}$ is calculated based on total site occupancies without considering orbital overlaps. Consequently, for intersite interactions, we have

$$\tilde{E} = \frac{1}{2}\sum_{I\neq J} V^{IJ}\left(\sum_{m\sigma} n^{I\sigma}_{mm}\right)\left(\sum_{m'\sigma'} n^{J\sigma'}_{m'm'}\right). \quad (6)$$

It can be shown that the derivative of the exact-counting energy $E$ [Eq. (5)] with respect to $n^{IJ\sigma}_{mm'}$ depends on orbital information, while the derivative of the effective counting energy $\tilde{E}$ with respect to $n^{IJ\sigma}_{mm'}$ or $n^{I\sigma}_{mm}$ is orbital-independent, as expected for an effective-field model.

Subtracting Eq. (6) from the exact counting expression in Eq. (5) yields the intersite Hubbard correction

$$E_V = -\frac{1}{2}\sum_{I\neq J} V^{IJ}\sum_{\sigma}\sum_{mm'} n^{IJ\sigma}_{mm'} n^{JI\sigma}_{m'm}. \quad (7)$$

Following the derivation of the onsite Hubbard correction in Ref. [32, 33], the total energy correction $E_{U+V}$ is

$$E_{U+V} = \frac{1}{2}\sum_{I}\sum_{\sigma,m,m'} U^I(\delta_{mm'} - n^{I\sigma}_{mm'})\, n^{I\sigma}_{m'm} - \frac{1}{2}\sum_{I\neq J}\sum_{\sigma,m,m'} V^{IJ} n^{IJ\sigma}_{mm'} n^{JI\sigma}_{m'm}. \quad (8)$$

In this work, both onsite and intersite Hubbard corrections are calculated from first principles as the second derivative of energy with respect to orbital occupations [33], using linear response formulation in the framework of DFPT [21, 22, 25, 32]. As both corrections aim to eliminate electron self-interaction, they yield more localized states compared to the DFT KS states. In the following section, we assess how such orbital localization would affect the frequency-dependent dielectric response and the resulting refractive indices of the materials of interest.

## Results and Discussion

### Influence of the selected projection manifold

Because the transition-metal–oxygen bonds in the noncentrosymmetric oxides considered here exhibit a pronounced covalent character, capturing the hybridization between their orbitals is essential for accurately predicting the electronic structure and the optical response. As shown in Fig. 2, the bare DFT overestimates refractive indices due to its notorious underestimation of bandgaps. Although onsite Hubbard $U$ corrections provide relatively accurate bandgaps, they suppress the optical anisotropy due to excessive electron localization. Therefore, the main achievement of DFT+$U$+$V$ is that it enables capturing the bond covalent characteristics and removes electron self-interaction errors without sacrificing anisotropy; in other words, inclusion of the intersite Hubbard $V$ term corrects the underestimated bandgap without suppressing the hybridization of transition metal (TM) and oxygen orbitals, which leads to accurate calculations of the refractive indices and optical anisotropy (Fig. 2). The Hubbard $U$+$V$ parameters calculated using DFPT are shown in Table 1. The choice of the orbital manifolds to which Hubbard corrections are applied is critical for optical anisotropy, since even within the DFT+$U$+$V$ framework, the anisotropy can still be suppressed if onsite Hubbard parameters on oxygen orbitals are included. In this context, we calculate refractive indices using both DFPT and ACBN0 with different sets of Hubbard parameters; namely, $\{U_{\mathrm{TM}}, U_{\mathrm{O}}\}$; $\{U_{\mathrm{TM}}, V_{\mathrm{TM-O}}\}$; $\{U_{\mathrm{TM}}, U_{\mathrm{O}}, V_{\mathrm{TM-O}}\}$. For brevity, we denote these parameter sets as I, II, III, respectively.

As shown in Table 1, the intersite Hubbard parameters obtained with DFPT using parameter set III exhibit nearly identical values for apical and equatorial oxygen atoms. This loss of distinction between crystallographically inequivalent oxygen sites is accompanied by a pronounced suppression of optical anisotropy. This behavior can be attributed to the limited susceptibility of oxygen orbitals to occupation changes under an applied local potential perturbation [23], resulting in a weak linear response and correspondingly large onsite Hubbard parameters. The resulting overlocalization of the oxygen states diminishes the anisotropic hybridization responsible for the optical anisotropy, even in the presence of intersite Hubbard interactions. Consequently, the Hubbard manifold $U_{TM}, V_{TM-O}$ (set II) provides the most accurate description of the optical anisotropy within the DFPT framework.

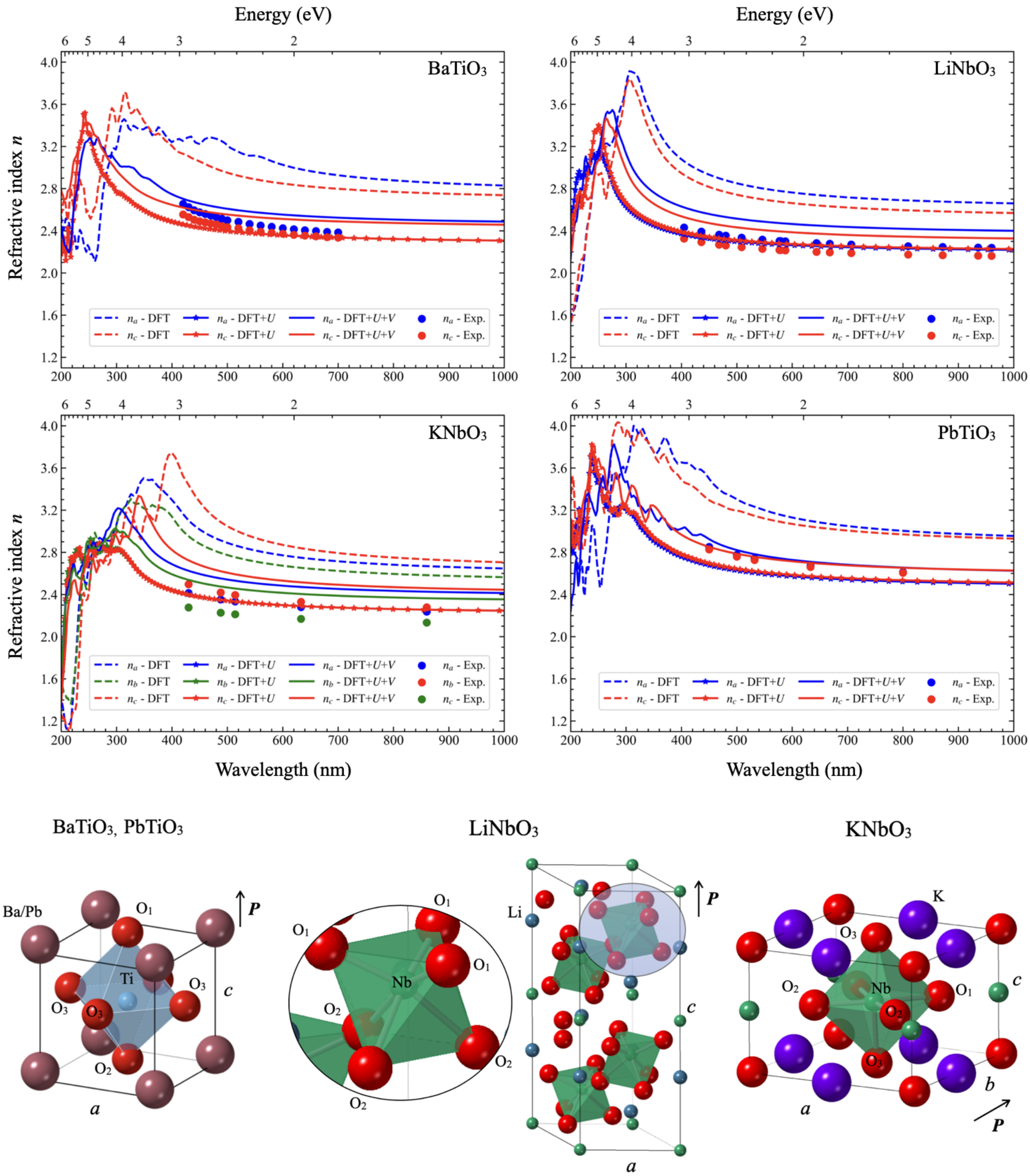

Figure 2: Refractive index dispersion in non-centrosymmetric oxides ($BaTiO_3$, $LiNbO_3$, $KNbO_3$, $PbTiO_3$) using DFT+$U$+$V$ against uncorrected DFT and DFT+$U$, showing optical anisotropy suppression when sole Hubbard $U$ is applied, experimental values as in [34–37]. The crystal structures with oxygen octahedra and spontaneous polarization vectors are shown for each case, with $BaTiO_3$ and $PbTiO_3$ in their tetragonal phase, $LiNbO_3$ in rhomohedral (conventional hexagonal cell is shown), and $KNbO_3$ in the orthorhombic phase. Oxygen atoms are labeled to indicate the calculated intersite Hubbard parameters as shown in Table 1, where $O_1$, $O_2$ represent the short-bond and long-bond oxygens in each octahedron, respectively, and $O_3$, are oxygens bonded to the transition metal perpendicular to the spontaneous polarization direction. The refractive indices $n_a$, $n_b$, $n_c$ correspond to the $\boldsymbol{a}$, $\boldsymbol{b}$, $\boldsymbol{c}$ directions, respectively as indicated by lattice parameters. For DFT+$U$+$V$, onsite Hubbard parameters on Ti, Nb and intersite parameters Ti–O, Nb–O are included, while onsite parameters on O are excluded.

Table 1: Calculated Hubbard parameters for the non-centrosymmetric oxides $BaTiO_3$, $LiNbO_3$, $KNbO_3$, and $PbTiO_3$ obtained using DFPT. Parameter sets I, II, and III correspond to $\{U_{TM}, U_O\}$, $\{U_{TM}, V_{TM-O}\}$, and $\{U_{TM}, U_O, V_{TM-O}\}$, respectively.

| | $BaTiO_3$ | | | $LiNbO_3$ | | | $KNbO_3$ | | | $PbTiO_3$ | | |
|---|---|---|---|---|---|---|---|---|---|---|---|---|
| Hubbard set | I | II | III | I | II | III | I | II | III | I | II | III |
| $U_{TM}$ (eV) | 4.73 | 5.15 | 4.88 | 3.30 | 3.45 | 3.33 | 3.46 | 3.58 | 3.48 | 4.83 | 5.02 | 4.87 |
| $U_O$ (eV) | 8.52 | | 8.66 | 7.90 | | 8.00 | 7.72 | | 7.83 | 8.10 | | 8.19 |
| $V_{TM-O_1}$ (eV) | | 1.31 | 1.08 | | 0.95 | 0.83 | | 1.05 | 0.97 | | 1.15 | 0.95 |
| $V_{TM-O_2}$ (eV) | | 1.01 | 1.08 | | 0.97 | 0.88 | | 1.10 | 0.98 | | 0.43 | 0.78 |
| $V_{TM-O_3}$ (eV) | | 1.27 | 1.08 | | | | | 1.07 | 0.95 | | 1.02 | 0.90 |

**Assessment of optical anisotropy calculations**

Recently, Ref. [30] reported that ACBN0 with parameter set III, $\{U_{TM}, U_O, V_{TM-O}\}$, yields accurate lattice parameters and bandgaps. While our calculations reproduce these structural and electronic trends, the birefringence remains substantially underestimated, as shown in Fig. 3. To enable a direct comparison between the two self-consistent Hubbard approaches, we also performed ACBN0 calculations using parameter set II, excluding the onsite oxygen correction. Although removal of the onsite oxygen correction improves the predicted optical response relative to set III, ACBN0 systematically overestimates both ferroelectric lattice distortions and birefringence of $BaTiO_3$, $PbTiO_3$, and $LiNbO_3$, resulting in less accurate equilibrium geometries than those obtained with DFPT, as shown in Table 2.

These differences are consistent with the distinct theoretical foundations of the two approaches. As recently discussed in Ref. [42], DFPT and ACBN0 are not expected to produce identical Hubbard parameters because they are derived from different physical quantities. In ACBN0, the effective Hubbard interactions are obtained from renormalized Hartree–Fock Coulomb and exchange integrals that depend on the localized density matrix. In contrast, DFPT determines the Hubbard parameters from the second derivative of the DFT total energy with respect to localized orbital occupations under infinitesimal perturbations, thereby explicitly incorporating electronic screening through the linear-response formalism. Within the energy functional of Eq. (7), the DFPT definition of the intersite interaction naturally recovers the interpretation of $V$ as the difference between the exact and effective electron-counting contributions derived above. By contrast, because the ACBN0 interaction parameters are not obtained from the curvature of the ACBN0 total energy, no analogous correspondence is guaranteed. The distinction becomes particularly important in strongly covalent ferroelectric oxides, where long-range dielectric screening and electronic redistribution extend beyond the information contained in the instantaneous localized density matrix. Consequently, properties that depend simultaneously on electronic hybridization and structural distortions, such as optical anisotropy, exhibit larger discrepancies between DFPT and ACBN0 than quantities governed primarily by lo-cal correlation effects. While ACBN0 has been shown to perform well for transition-metal oxides with predominantly local electronic correlations [42], the present results indicate that DFPT provides a more consistent description of the optical anisotropy in the present class of ferroelectric oxides.

Table 2: Relaxed lattice parameters for the non-centrosymmetric oxides $BaTiO_3$, $LiNbO_3$, $KNbO_3$, and $PbTiO_3$ obtained using DFPT and ACBN0. Parameter sets I, II, and III correspond to $\{U_{TM}, U_O\}$, $\{U_{TM}, V_{TM-O}\}$, and $\{U_{TM}, U_O, V_{TM-O}\}$, respectively. Experimental values are from Ref. [38–41]. ACBN0 sets I, III for $BaTiO_3$, $LiNbO_3$, $PbTiO_3$ are similar to those in Ref. [30]. The oxygen-atom labeling is shown in Fig. 2. Positive off-centering values are along the spontaneous polarization direction.

| | DFPT | | | ACBN0 | | | Exp. |
|---|---|---|---|---|---|---|---|
| Hubbard set | I | II | III | I | II | III | |
| **$BaTiO_3$** | | | | | | | |
| $a$ (Å) | 3.978 | 3.966 | 3.975 | 3.970 | 3.937 | 3.945 | 3.99 |
| $c/a$ | 1.000 | 1.030 | 1.000 | 1.000 | 1.064 | 1.034 | 1.01 |
| $\Delta z_{Ti}$ (Å) | 0 | 0.067 | 0.001 | 0 | 0.088 | 0.067 | 0.06 |
| $\Delta z_{O_1}$ (Å) | 0 | –0.145 | –0.003 | 0 | –0.244 | –0.173 | –0.093 |
| $\Delta z_{O_3}$ (Å) | 0 | –0.090 | –0.002 | 0 | –0.148 | –0.107 | –0.056 |
| **$LiNbO_3$** | | | | | | | |
| $a$ (Å) | 5.104 | 5.131 | 5.105 | 5.10 | 5.144 | 5.113 | 5.151 |
| $c/a$ | 2.690 | 2.697 | 2.689 | 2.69 | 2.732 | 2.698 | 2.692 |
| $\Delta z_{Nb}$ (Å) | –0.447 | –0.446 | –0.451 | –0.454 | –0.418 | –0.441 | –0.456 |
| $\Delta z_{O_1}$ (Å) | –0.609 | –0.713 | –0.621 | –0.641 | –0.769 | –0.713 | –0.722 |
| **$KNbO_3$** | | | | | | | |
| $a$ (Å) | 5.621 | 5.689 | 5.621 | 5.611 | 5.792 | 5.676 | 5.695 |
| $b$ (Å) | 5.621 | 5.708 | 5.621 | 5.611 | 5.869 | 5.696 | 5.723 |
| $c$ (Å) | 3.974 | 3.969 | 3.973 | 3.967 | 3.918 | 3.949 | 3.974 |
| $\Delta y_{Nb}$ (Å) | 0 | 0.067 | 0.010 | 0 | 0.110 | 0.064 | 0.070 |
| $\Delta y_{O_1}$ (Å) | 0 | –0.105 | –0.009 | 0 | –0.175 | –0.122 | –0.125 |
| $\Delta y_{O_3}$ (Å) | 0 | –0.117 | –0.010 | 0 | –0.223 | –0.140 | –0.138 |
| **$PbTiO_3$** | | | | | | | |
| $a$ (Å) | 3.932 | 3.850 | 3.923 | 3.90 | 3.819 | 3.799 | 3.904 |
| $c/a$ | 1.003 | 1.150 | 1.016 | 1.02 | 1.204 | 1.224 | 1.063 |
| $\Delta z_{Ti}$ (Å) | –0.118 | –0.207 | –0.142 | –0.143 | –0.252 | –0.267 | –0.170 |
| $\Delta z_{O_1}$ (Å) | –0.172 | –0.671 | –0.271 | –0.306 | –0.836 | –0.867 | –0.456 |
| $\Delta z_{O_3}$ (Å) | –0.218 | –0.656 | –0.318 | –0.352 | –0.750 | –0.775 | –0.456 |

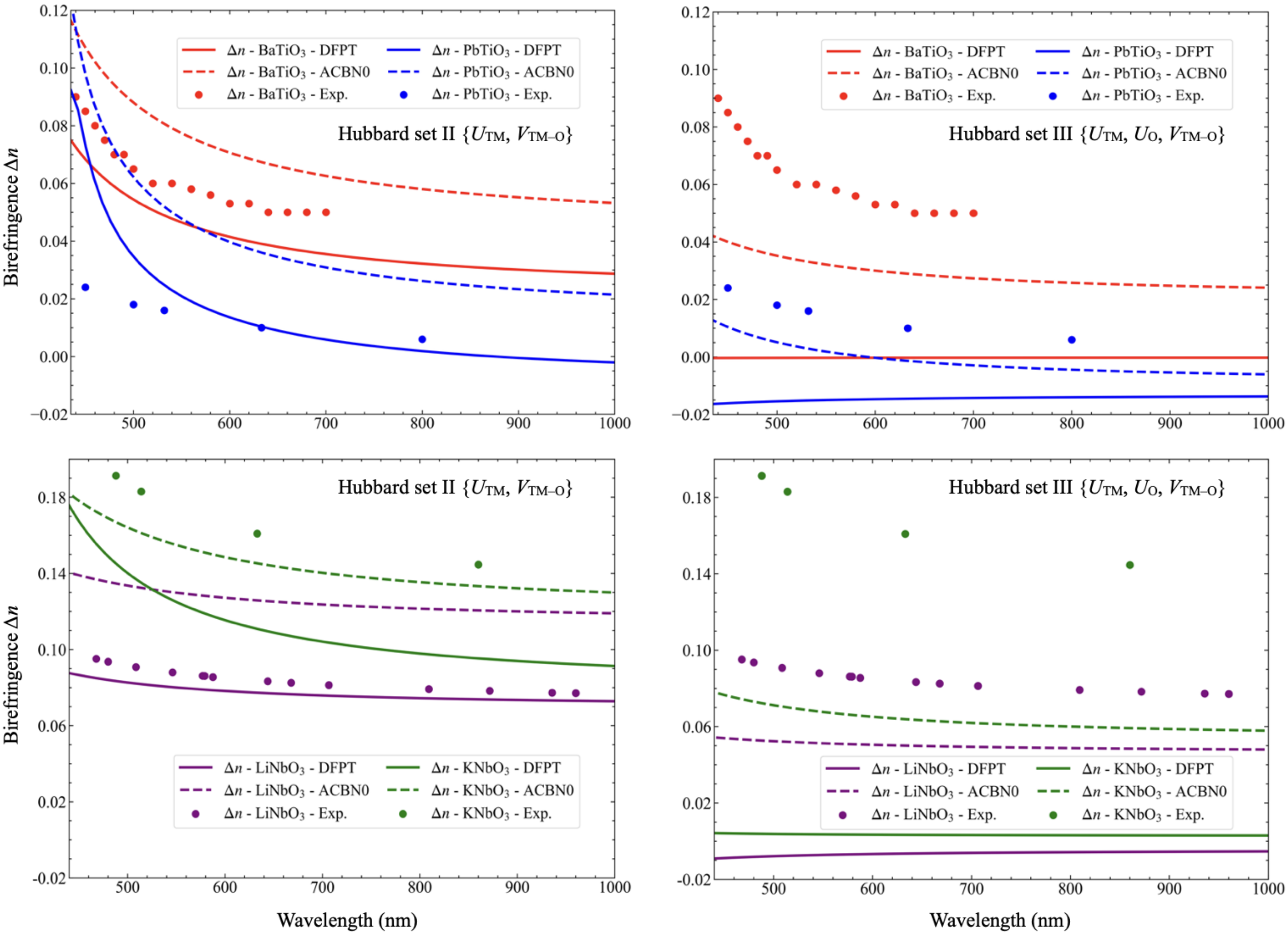

Figure 3: Birefringence dispersion of non-centrosymmetric oxides ($BaTiO_3$, $LiNbO_3$, $KNbO_3$, $PbTiO_3$) calculated using DFPT and ACBN0 for different sets of Hubbard parameters. Experimental values as from Refs. [34–37]. DFPT using Hubbard set II gives more consistent description of birefringence measurements.

To assess the impact of these methodological differences, we benchmarked the predicted structural and optical properties against the available experimental measurements. As shown in Table 2 and Fig. 3, the DFPT calculations employing the Hubbard set II, $\{U_{\mathrm{TM}}, V_{\mathrm{TM-O}}\}$, reproduce the experimental crystal geometries and optical anisotropies with good overall accuracy across the four benchmark ferroelectric oxides. For $BaTiO_3$, the calculated birefringence is in good agreement with the experimental measurements of [34] in the 400–700 nm wavelength range. For $PbTiO_3$, the DFPT results closely follow the Sellmeier fit reported in Ref. [37]; however, as noted in that work, the fit is not reliable below 500 nm. In contrast, the experimental measurements from Ref. [43] report an approximately fourfold increase in birefringence within this spectral region, consistent with our DFPT predictions. For $LiNbO_3$, the calculated birefringence shows excellent agreement with the measurements in Ref. [35]. $KNbO_3$ represents the only exception, for which ACBN0 yields birefringence values in closer agreement with experiment. This behavior is likely the result of compensating effects: the overestimated ferroelectric distortion enhances the birefringence, whereas the simultaneously overestimated bandgap suppresses it, leading to fortuitous agreement with experiment.

Table 3 summarizes the bandgaps obtained with both DFPT and ACBN0 for the different Hubbard parameter sets. The corresponding ACBN0 Hubbard parameters are provided in the Supplementary Information. Parameter sets I and III are consistent with those reported in [30]; however, the calculations were repeated here to evaluate the dielectric response and optical anisotropy. The ordinary and extraordinary refractive index dispersion for ACBN0 sets II and III are provided in the Supplementary Information (SI).

Table 3: Calculated bandgaps (direct/indirect) for the non-centrosymmetric oxides $BaTiO_3$, $LiNbO_3$, $KNbO_3$, and $PbTiO_3$ obtained using DFPT, ACBN0 vs GW. Parameter sets I, II, and III correspond to $\{U_{\mathrm{TM}}, U_{\mathrm{O}}\}$, $\{U_{\mathrm{TM}}, V_{\mathrm{TM-O}}\}$, and $\{U_{\mathrm{TM}}, U_{\mathrm{O}}, V_{\mathrm{TM-O}}\}$, respectively. ACBN0 sets I, III for $BaTiO_3$, $LiNbO_3$, $PbTiO_3$ are similar to those used in Ref. [30].

| | DFPT $\epsilon_g$ (eV) | | | ACBN0 $\epsilon_g$ (eV) | | | GW+BSE $\epsilon_g$ (eV) | Exp. $\epsilon_g$ (eV) |
|---|---|---|---|---|---|---|---|---|
| Hubbard set | I | II | III | I | II | III | | |
| $BaTiO_3$ | 3.66/3.66 | 3.34/2.85 | 4.02/4.02 | 2.74/2.74 | 3.33/2.70 | 4.01/3.68 | 4.40/3.90 [11] | 3.27/– [44] |
| $LiNbO_3$ | 4.56/4.52 | 4.13/4.07 | 4.82/4.79 | 4.28/4.26 | 5.25/4.95 | 5.61/5.58 | –/4.71 [17] | 4.12/3.95 [45] |
| $KNbO_3$ | 3.25/3.21 | 2.66/2.51 | 3.46/3.41 | 2.82/2.78 | 3.87/3.51 | 4.03/3.84 | 3.99/3.13 [18] | 3.28/– [46] |
| $PbTiO_3$ | 2.79/2.72 | 2.99/2.57 | 3.27/2.98 | 2.10/1.83 | 2.98/2.73 | 3.96/2.95 | 3.49/3.20 [19] | 3.10/– [47], 3.60/– [48] |

## Birefringent phase-matching landscapes

Motivated by the overall agreement between DFPT predictions and experiment, we employ this approach to construct phase-matching landscapes for the investigated materials (Fig. 4). These landscapes are obtained by mapping the frequency-scaled refractive-index mismatch

$$\Delta n(\omega, \gamma) = n_{\mathrm{o}}(\omega) - n_{\mathrm{e}}(\gamma\omega), \tag{9}$$

where the scaling factor $\gamma$ is treated as a continuous variable. Here, $n_{\mathrm{o}}$ and $n_{\mathrm{e}}$ denote the ordinary and extraordinary refractive indices for the uniaxial crystals $BaTiO_3$, $LiNbO_3$, and $PbTiO_3$. For the biaxial crystal $KNbO_3$, they are replaced by the refractive indices along the selected principal axes, taken here as the largest and smallest principal refractive indices. Integer values of $\gamma$ correspond to harmonic-generation processes, whereas non-integer values represent arbitrary frequency-conversion pairs. This representation identifies the loci satisfying the phase-matching condition together with the surrounding regions over which phase matching can be maintained, thereby providing a convenient framework for analyzing phase-matching bandwidths, designing tunable nonlinear optical devices, and exploring novel frequency-conversion schemes. As shown in Fig. 4, $LiNbO_3$ satisfies the phase-matching condition for SHG at a fundamental photon energy of approximately 1.1 eV, in agreement with the experimentally demonstrated generation of green light. In contrast, the SHG phase-matching condition in $BaTiO_3$ occurs in the near-infrared (NIR) spectral region. $KNbO_3$ exhibits a comparatively flatter phase-matching landscape owing to its larger birefringence, whereas the strong chromatic dispersion of $PbTiO_3$ dominates over its birefringence, preventing birefringent phase matching for conventional near-infrared and visible laser wavelengths.

To quantitatively compare phase-matching landscapes across different materials, we introduce two metrics: the phase-matching fraction $\Gamma$ (that measures the size of the phase-matching region) and the renormalized Dirichlet energy $\mathcal{E}$ [49](that measures the gradient of $\Delta n$ around the phase-matching region). These adimensional phase-matching metrics are expressed as follows:

$$\Gamma[f] = \Omega^{-1} \int_{\Omega} \delta_\sigma(f(\boldsymbol{\omega}))\mathrm{d}^2\boldsymbol{\omega} \tag{10}$$

$$\mathcal{E}[f] = \Gamma^{-1}[f] \int_{\Omega} |\nabla f(\boldsymbol{\omega})|^2 \delta_\sigma(f(\boldsymbol{\omega}))\mathrm{d}^2\boldsymbol{\omega}, \tag{11}$$

where $\boldsymbol{\omega} = (\omega, \gamma\omega) \equiv (\nu\cos(\theta), \nu\sin(\theta))$ and $\mathrm{d}^2\boldsymbol{\omega} = \nu \mathrm{d}\nu \mathrm{d}\theta$ (by applying a polar-coordinate transformation), $\Omega = \int \mathrm{d}^2\boldsymbol{\omega}$, and $f(\boldsymbol{\omega}) \equiv \Delta n(\omega, \gamma)$ represents the refractive-index mismatch function. Since only regions in the vicinity of $f = 0$ contribute to practical phase matching, a direct comparison of global landscape gradients is not physically meaningful. Therefore, we incorporate the weighting function $\delta_\sigma(f) = (2\pi\sigma^2)^{-1}\exp(-f^2/(2\sigma^2))$, with a broadening parameter $\sigma$, to restrict the evaluation to regions close to the phase-matching condition. A low renormalized Dirichlet energy indicates a flatter phase-matching landscape and thus a broader bandwidth, provided that the phase-matching condition is satisfied in the frequency domain of interest.

As a validation, we calculated $\Gamma$ and $\mathcal{E}$ for the SHG phase-matching landscape, $\Delta n(\omega, \gamma = 2)$ of $BaTiO_3$, $LiNbO_3$, $KNbO_3$, and $PbTiO_3$ for $\omega$ ranging from 0.8 to 2 eV, as shown in Table 4. We found that $LiNbO_3$ exhibits the lowest $\mathcal{E}$, indicating the most favorable phase-matching landscape among the considered materials. With $\sigma = 5 \times 10^{-3}$, the calculated $\mathcal{E}$ is $9.1 \times 10^{-3}$.

Table 4: Calculated phase-matching fraction $\Gamma$ and renormalized Dirichlet energy $\mathcal{E}$ for the SHG phase-matching landscape, $\Delta n(\omega, \gamma = 2)$ of $BaTiO_3$, $LiNbO_3$, $KNbO_3$, and $PbTiO_3$ for $\omega$ ranging from 0.8 to 2 eV.

| | $\Gamma$ | $\mathcal{E} \times 10^{-3}$ |
|---|---|---|
| $BaTiO_3$ | $\approx 0$ | |
| $LiNbO_3$ | 4.1 | 9.1 |
| $KNbO_3$ | 3.8 | 13.5 |
| $PbTiO_3$ | $\approx 0$ | |

By contrast, $KNbO_3$ shows $\mathcal{E} = 13.5 \times 10^{-3}$, indicating a steeper landscape close to its phase-matching condition. The remaining materials exhibit vanishing phase-matching fractions $\Gamma$ over the studied frequency domain. Beyond benchmarking different materials, this dimensionless descriptor provides a scalar measure of the phase-matching performance of optical materials that can be incorporated into screening workflows.

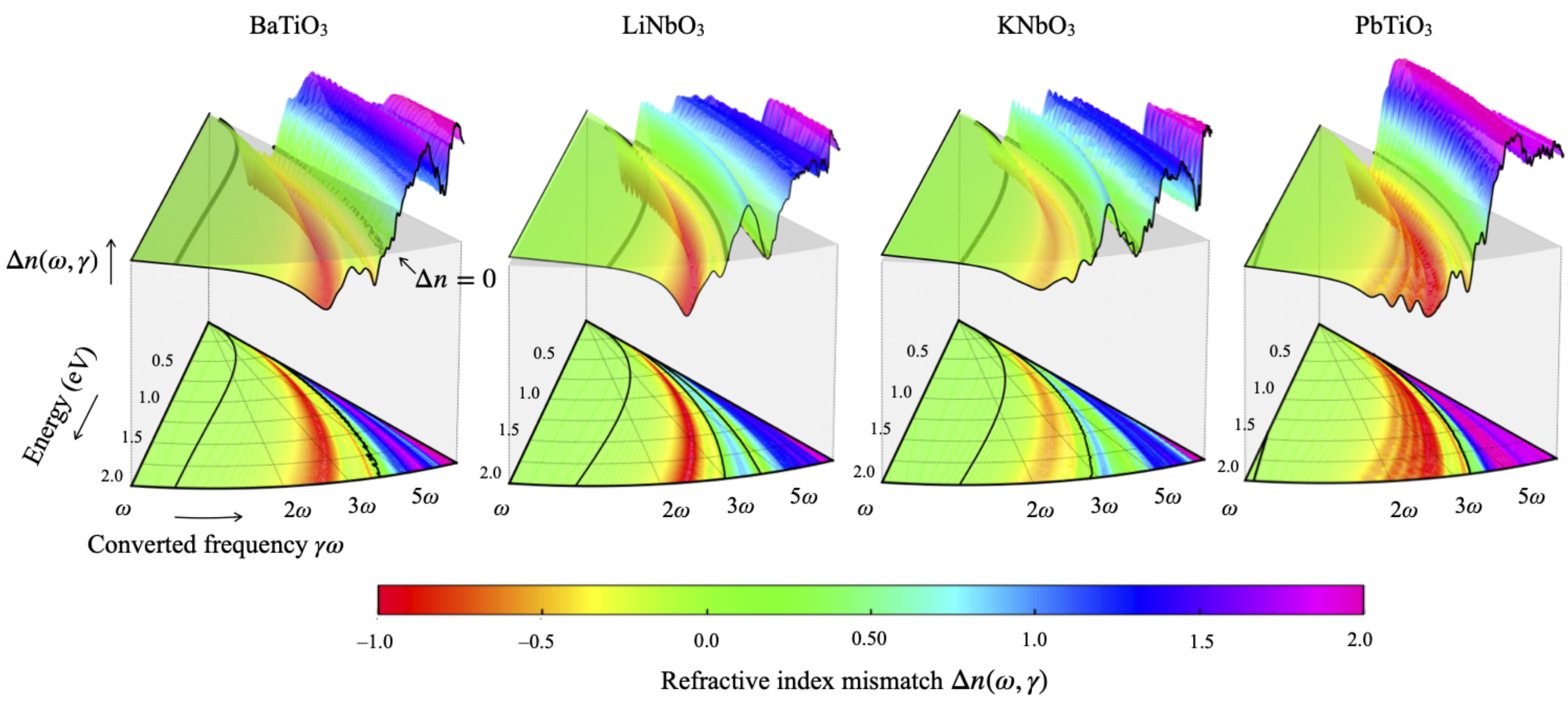

Figure 4: Birefringent phase-matching landscapes of $BaTiO_3$, $LiNbO_3$, $KNbO_3$, and $PbTiO_3$, and their polar projections. The radial and angular coordinates represent the pump energy and generated harmonics, respectively, while the vertical axis corresponds to the frequency-scaled refractive-index mismatch, $\Delta n(\omega, \gamma)$. The horizontal gray plane denotes the phase-matching condition, $\Delta n(\omega, \gamma) = 0$, whereas the corresponding phase-matching trajectories and their polar projections are shown in black.

### Optical response of $BaTiO_3$ solid solutions

Having established that the DFPT-based DFT+$U$+$V$ approach with the Hubbard set II, $\{U_{\mathrm{TM}}, V_{\mathrm{TM-O}}\}$, provides a consistent description of the anisotropic optical response of bench-marked ferroelectric oxides, we next demonstrate its applicability to the computational design of optical materials. Because of its significantly lower computational cost than that of higher-level electronic-structure methods, the approach is well suited for high-throughput screening of chemically substituted ferroelectrics and their solid solutions. As a representative case study, we investigate the effect of chemical substitution on the optical properties of $BaTiO_3$. Two compositions are considered: A-site substitution with Zn, and Bi-Mn co-substitution at the A and B sites, respectively. The corresponding phase-matching landscapes are compared with $BaTiO_3$ and $LiNbO_3$ to elucidate doping effects on optical anisotropy. While numerous first-principles studies examined the influence of doping on the electronic and optical properties of $BaTiO_3$ [50–52], the resulting anisotropic optical response has been much less studied. The self-consistent Hubbard parameters are summarized in Table 5 and the projected density of states (PDOS) of the parent compound and the substituted cases are shown in (Fig. 5).

Our calculations show that the relaxed structure approaches an ideal tetragonal distortion for both Zn substitution and Bi–Mn co-substitution cases with an increased tetragonality ratio relative to pristine $BaTiO_3$, in agreement with available experimental observations [53] for the Bi–Mn substituted case. The enhanced structural distortion is accompanied by a substantial increase in the optical anisotropy for Bi–Mn co-substitution (Fig. 6). As shown in (Fig. 5), the electronic structure of $Ba_7BiTi_7MnO_{24}$ reveals that Mn introduces mid-gap states through hybridization between the Mn-3$d$ and O-2$p$ orbitals. Because these mid-gap states are dominated by directional orbitals $d_{x^2-y^2}$, the resulting optical transitions become polarization-dependent, producing anisotropic optical bandgap narrowing.

In contrast to Zn substitution, which produces only minor changes in optical anisotropy, the simultaneous incorporation of a large A-site cation possessing stereochemically active lone-pair electrons (Bi) and a smaller B-site cation (Mn) markedly reconstructs the local octahedral environment, predominantly along the polar axis. This reconstruction is characterized by a pronounced elongation of the shorter apical Mn–O bond, while the longer apical and equatorial bonds remain nearly unchanged, indicating that the structural relaxation is strongly anisotropic. This behavior originates from dopant-induced changes in the local electrostatic environment together with Mn–O hybridization. As a result, the anisotropic optical response is significantly strengthened, yielding an almost fourfold increase in birefringence relative to pristine $BaTiO_3$, as illustrated by the cross-sections of the phase-matching landscape (Fig. 6). Consequently, the SHG phase-matching frequency of $Ba_7BiTi_7MnO_{24}$ is blue-shifted relative to its parent compound, enabling birefringent Type-I SHG phase matching over a broader portion of the mid- to near-infrared spectral range (0.75-1.0 eV). Here, Type-I SHG corresponds to the interaction of two ordinarily polarized fundamental waves that generate an extraordinarily polarized second-harmonic wave. This improvement is achieved without significantly compromising the optical transparency window in the same energy range (Fig. 7a). More importantly, the pronounced absorption anisotropy of $Ba_7BiTi_7MnO_{24}$ results in a large dichroic ratio in the visible spectral range (Fig. 7), suggesting its potential for polarization-sensitive photodetectors and integrated photonic applications.

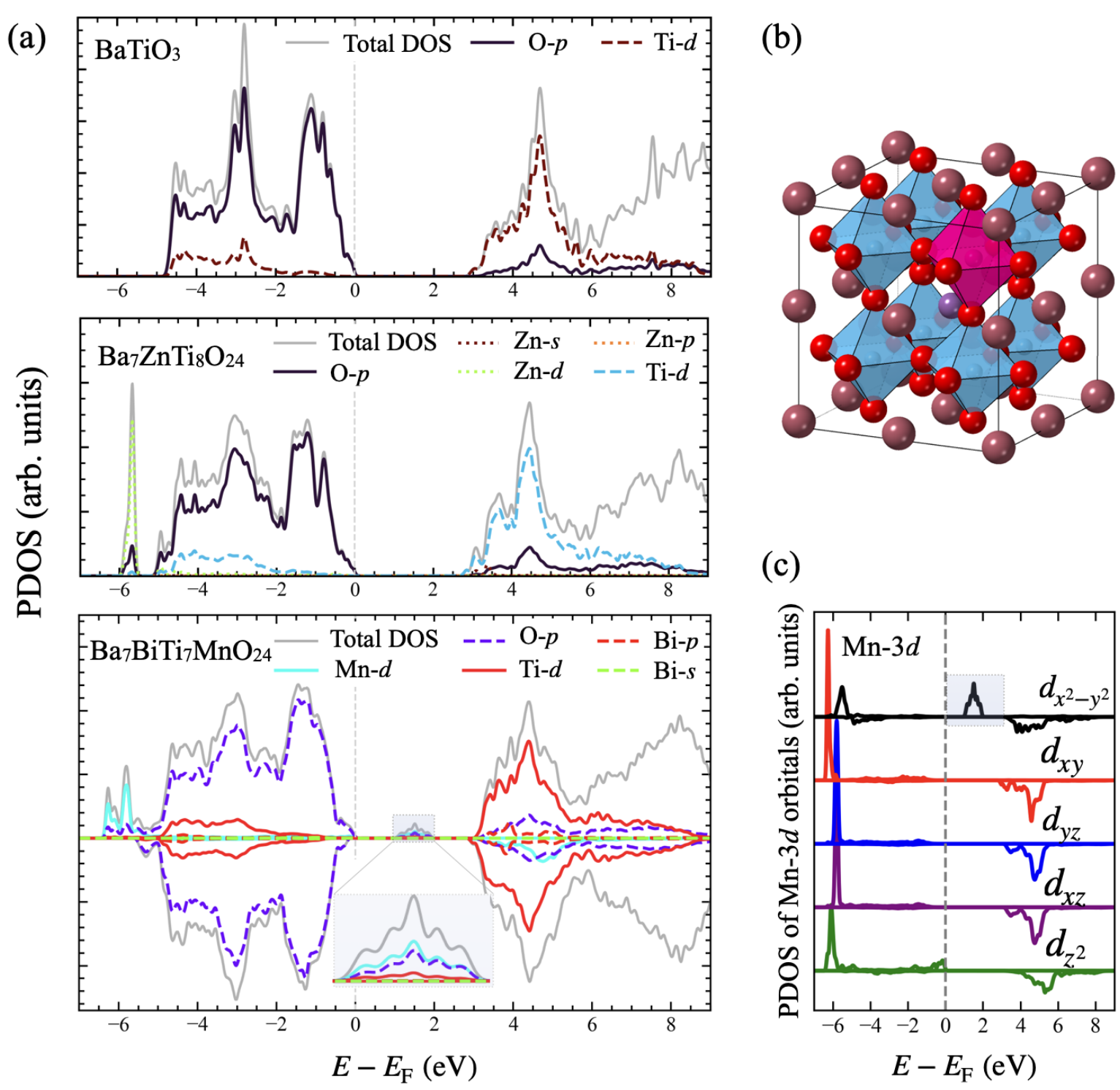

Figure 5: Density of states of cation-substituted $BaTiO_3$. The energy scale is referenced to the Fermi level, $E_F$, such that $E - E_F = 0$ eV corresponds to the Fermi energy; negative values denote valence states and positive values denote conduction states. (a) Projected density of states (PDOS) of the parent compound, $Ba_7ZnTi_8O_{24}$, and $Ba_7BiTi_7MnO_{24}$. PDOS decomposes the total electronic density of states into contributions from selected atomic species and orbitals, thereby revealing the orbital character of the electronic bands. The inset for $Ba_7BiTi_7MnO_{24}$ shows the mid-gap energy states induced by Mn-3$d$ and O-2$p$ orbital hybridization. (b) Atomic structure of $Ba_7BiTi_7MnO_{24}$. A 2×2×2 supercell containing 40 atoms is shown. (c) PDOS for Mn-3$d$ orbitals in $Ba_7BiTi_7MnO_{24}$.

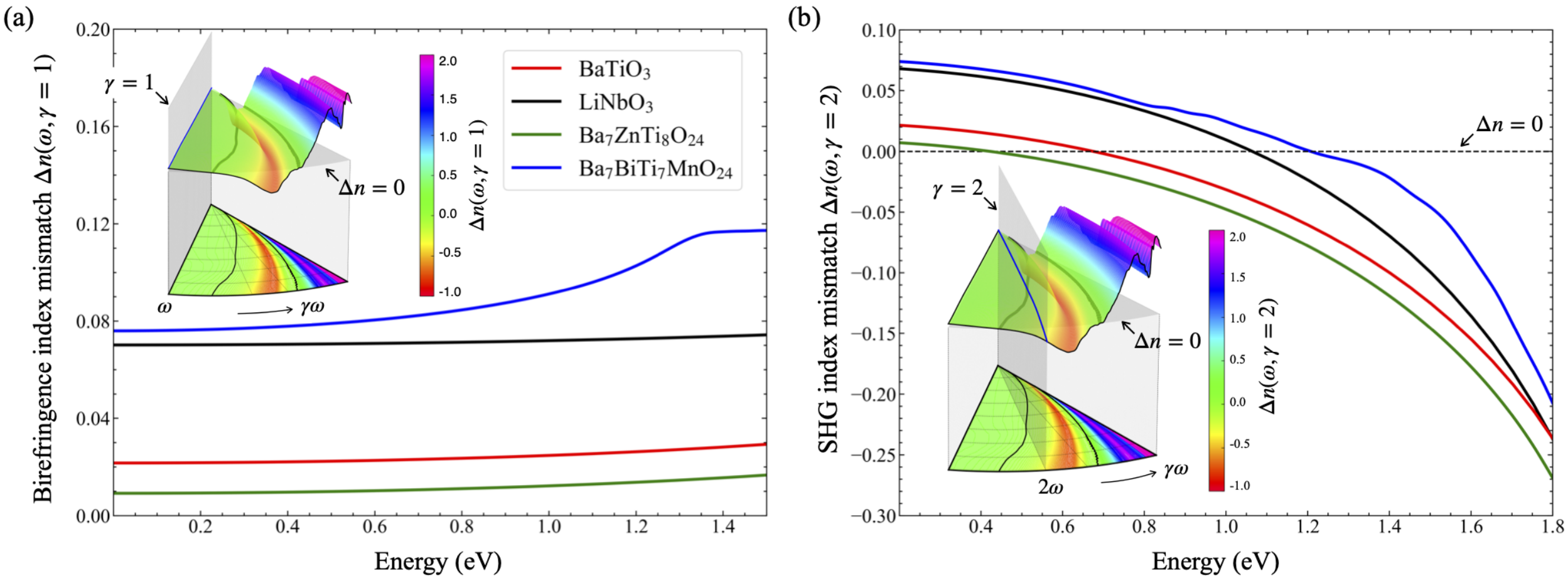

Figure 6: (a) Birefringence dispersion and (b) SHG cross section of $Ba_7ZnTi_8O_{24}$ and $Ba_7BiTi_7MnO_{24}$ compared to those of $LiNbO_3$ and pristine $BaTiO_3$. The 3D phase-matching landscape of the $Ba_7BiTi_7MnO_{24}$ case is shown in the inset, where the birefringence section and the SHG section corresponds to $\gamma = 1$ and $\gamma = 2$, respectively. The horizontal gray plane in the inset denotes the phase-matching condition, $\Delta n(\omega, \gamma) = 0$, whereas the corresponding phase-matching trajectories and their polar projections are shown in black. The SHG phase-matching locus in (b) represents $\Delta n(\omega, \gamma = 2) = 0$.

Table 5: Calculated non-empirical Hubbard parameters for $Ba_7ZnTi_8O_{24}$, and $Ba_7BiTi_7MnO_{24}$. $O_1$, $O_2$ label the apical oxygen sites, whereas $O_3$ and $O_{3'}$ label the equatorial oxygen sites rendered inequivalent due to the symmetry breaking induced by A-site cation substitution.

| | $Ba_7ZnTi_8O_{24}$ | $Ba_7BiTi_7MnO_{24}$ |
|---|---|---|
| $U_{\mathrm{Ti}}$ (eV) | 4.80 | 4.94 |
| $V_{\mathrm{Ti-O_1}}$ (eV) | 1.14 | 1.24 |
| $V_{\mathrm{Ti-O_2}}$ (eV) | 0.74 | 0.80 |
| $V_{\mathrm{Ti-O_3}}$ (eV) | 1.11 | 1.26 |
| $V_{\mathrm{Ti-O_3'}}$ (eV) | 0.98 | 0.97 |
| $U_{\mathrm{Mn}}$ (eV) | | 5.73 |
| $V_{\mathrm{Mn-O_1}}$ (eV) | | 0.98 |
| $V_{\mathrm{Mn-O_2}}$ (eV) | | 0.64 |
| $V_{\mathrm{Mn-O_3}}$ (eV) | | 1.12 |
| $V_{\mathrm{Mn-O_3'}}$ (eV) | | 0.88 |

Overall, this study has shown that DFT+$U$+$V$ approach, with self-consistent Hubbard parameters calculated using DFPT, provides a computationally efficient framework for predicting the anisotropic optical response of non-centrosymmetric perovskite oxides. Our results demonstrate that the choice of Hubbard manifold is critical, as an inappropriate manifold can suppress optical anisotropy even within the DFT+$U$+$V$ formalism. With the appropriate manifold, DFPT reproduces the measured anisotropic optical properties across the benchmark compounds while retaining the efficiency needed for large-scale simulations. Because the Hubbard parameters are determined self-consistently from the local chemical environment, the same framework naturally extends to solid solutions without empirical parameter fitting. Applying this approach to Bi–Mn substituted $BaTiO_3$, we show that co-substitution induces polarization-dependent optical bandgap narrowing through mid-gap states, enhancing the visible-range dichroic ratio and blue-shifting the Type-I SHG phase-matching frequency. These results highlight the potential of self-consistent DFT+$U$+$V$ as a predictive tool for the computational discovery and design of optical materials.

## Computational Methods

Density-functional calculations were performed using Quantum ESPRESSO [54]. The plane-wave pseudopotential method is implemented, where we used the generalized-gradient approximation for the exchange-correlation functional, with the revised Perdew-Burke-Ernzerhof parameterization for solids (PBEsol) [55, 56]. Scalar-relativistic norm-conserving pseudopotentials were obtained from the PseudoDojo database (version 0.4, standard accuracy set) [57], generated using the ONCVPSP scheme [58]. Kinetic energy cutoff of 120 Ry was selected. The Brillouin-zone sampling for computing electronic ground states used a Monkhorst–Pack centered 8×8×8 $k$-point mesh. After performing full geometry optimization at the DFT level of theory, the calculations of Hubbard parameters $U$ and $V$ were carried out using DFPT as implemented in the `hp` code [25], which is part of Quantum ESPRESSO. Hubbard parameters for ACBN0 method are obtained using Löwdin orthogonalized atomic orbitals as implemented in Qunatum ESPRESSO [27–29]. The dielectric response in momentum space is calculated using the independent-particle approximation as a function of frequency:

$$\varepsilon_{ij} = \varepsilon_0\delta_{ij} - \frac{e^2\hbar^2}{\Omega m_{\mathrm{e}}^2}\sum_{n,m,m\neq n}\frac{1}{(E_n-E_m)^2}\times \left(\frac{\langle\varphi_m|\hat{P}_i|\varphi_n\rangle\langle\varphi_n|\hat{P}_j|\varphi_m\rangle}{E_n-E_m+\hbar\omega+\mathrm{i}\hbar\Gamma}+\frac{\langle\varphi_n|\hat{P}_i|\varphi_m\rangle\langle\varphi_m|\hat{P}_j|\varphi_n\rangle}{E_n-E_m-\hbar\omega-\mathrm{i}\hbar\Gamma}\right),$$

where the double indexing reveals the tensorial nature of $\boldsymbol{\varepsilon}(\omega)$, $\hat{\boldsymbol{P}}$ is the polarization operator, $|\varphi_n\rangle$ are KS states, $\Gamma$ is the damping parameter, $\Omega$ is the unit cell volume, and $\varepsilon_0$ is the permittivity of the free space. The summation index runs over band transitions ($n$, $m$ are band indices; $n \neq m$). The dielectric response can be expressed in terms of its real and imaginary parts as

$$\varepsilon(\omega) = \varepsilon_1(\omega) + i\varepsilon_2(\omega). \quad (12)$$

Consequently, the refractive index $n(\omega)$ and the extinction coefficient $K(\omega)$ can be calculated as

$$n(\omega) = \frac{1}{\sqrt{2\varepsilon_0}}\left(\sqrt{\varepsilon_1^2(\omega)+\varepsilon_2^2(\omega)}+\varepsilon_1(\omega)\right)^{\frac{1}{2}} \quad (13)$$

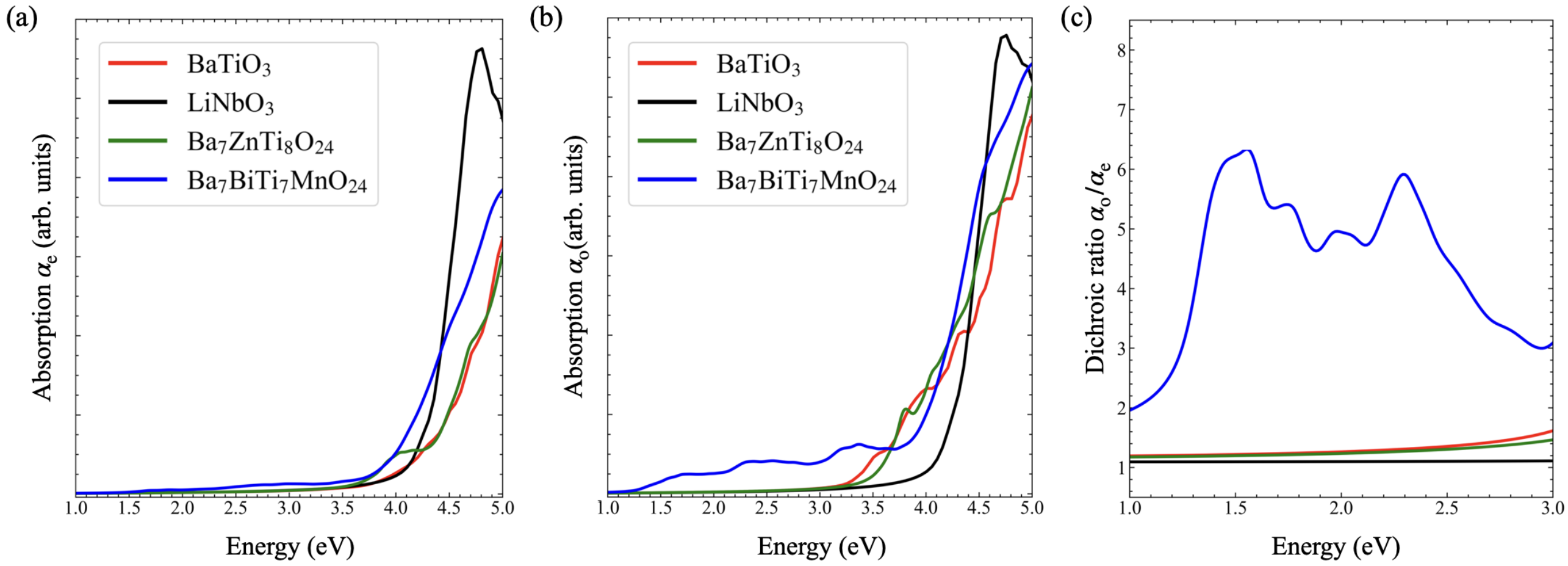

Figure 7: Absorption spectra in the extraordinary (a) and ordinary (b) directions, and dichroic ratio (c) of $Ba_7ZnTi_8O_{24}$, and $Ba_7BiTi_7MnO_{24}$ vs $LiNbO_3$ and pristine $BaTiO_3$.

$$K(\omega) = \frac{1}{\sqrt{2\varepsilon_0}}\left(\sqrt{\varepsilon_1^2(\omega) + \varepsilon_2^2(\omega)} - \varepsilon_1(\omega)\right)^{\frac{1}{2}}. \quad (14)$$

The cation-substituted structures were modeled using 40-atom (2×2×2) tetragonal supercells. Structural relaxations were performed until the residual forces on all atoms were below 0.01 eV/Å, and the Brillouin zone was sampled using a Monkhorst–Pack mesh corresponding to a reciprocal-space sampling density of 0.25 $Å^{-1}$. Spin-polarized calculations were carried out for the Bi–Mn co-substituted system. The projected density of states (PDOS) was evaluated using Löwdin-orthogonalized atomic orbitals.

## Data availability

The data that support the findings of this study are available from the corresponding author on a reasonable request.

## Author contributions

M.S.M.M.A.: Conceptualization; First-principles calculations; Formal analysis; Writing – original draft; Writing – review and editing. I.D.: Conceptualization; Methodology; Formal analysis; Supervision; Writing – review and editing; Funding acquisition.

## Competing interests

The authors declare no competing interests.

## Acknowledgment

This work was primarily supported by the Center for Computational Mesoscale Materials Science (COMMS), as a part of the Computational Materials Sciences program of the US Department of Energy, Office of Science, Basic Energy Sciences, under award number DE-SC0020145. The first-principles simulations in this work were performed using Anvil supercomputer at Purdue University and Bridges-2 at Pittsburgh Supercomputing Center (PSC) through the allocation MAT250027 from the Advanced Cyberinfrastructure Coordination Ecosystem: Services & Support (ACCESS) program, which is supported by U.S. National Science Foundation grants #2138259, #2138286, #2138307, #2137603, and #2138296. The authors gratefully acknowledge insightful discussions with Long-Qing Chen, Aiden Ross and Venkatraman Gopalan at the Pennsylvania State University during the early stages of this work.

## Supplementary Information

Table S1: Calculated Hubbard parameters for the non-centrosymmetric oxides $BaTiO_3$, $LiNbO_3$, $KNbO_3$, and $PbTiO_3$ obtained using ACBN0. Parameter sets I, II, and III correspond to $\{U_{TM}, U_O\}$, $\{U_{TM}, V_{TM-O}\}$, and $\{U_{TM}, U_O, V_{TM-O}\}$, respectively. ACBN0 sets I, III for $BaTiO_3$, $LiNbO_3$, $PbTiO_3$ are similar to those in Ref. [30].

| | $BaTiO_3$ | | | $LiNbO_3$ | | | $KNbO_3$ | | | $PbTiO_3$ | | |
|---|---|---|---|---|---|---|---|---|---|---|---|---|
| Hubbard set | I | II | III | I | II | III | I | II | III | I | II | III |
| $U_{TM}$ (eV) | 0.24 | 0.41 | 0.25 | 0.29 | 0.52 | 0.37 | 0.31 | 0.55 | 0.37 | 0.17 | 0.30 | 0.21 |
| $U_{O_{1,2}}$ (eV) | 7.83 | | 7.36 | 8.47 | | 8.30 | 8.20 | | 7.91 | 7.44 | | 7.12 |
| $U_{O_3}$ (eV) | 7.83 | | 7.72 | | | | 8.19 | | 7.99 | 7.33 | | 7.11 |
| $V_{TM-O_1}$ (eV) | | 2.82 | 2.52 | | 2.86 | 2.70 | | 2.74 | 2.48 | | 2.65 | 2.59 |
| $V_{TM-O_2}$ (eV) | | 2.11 | 2.04 | | 2.47 | 2.44 | | 2.27 | 2.20 | | 1.63 | 1.59 |
| $V_{TM-O_3}$ (eV) | | 2.59 | 2.35 | | | | | 2.59 | 2.37 | | 2.34 | 2.30 |

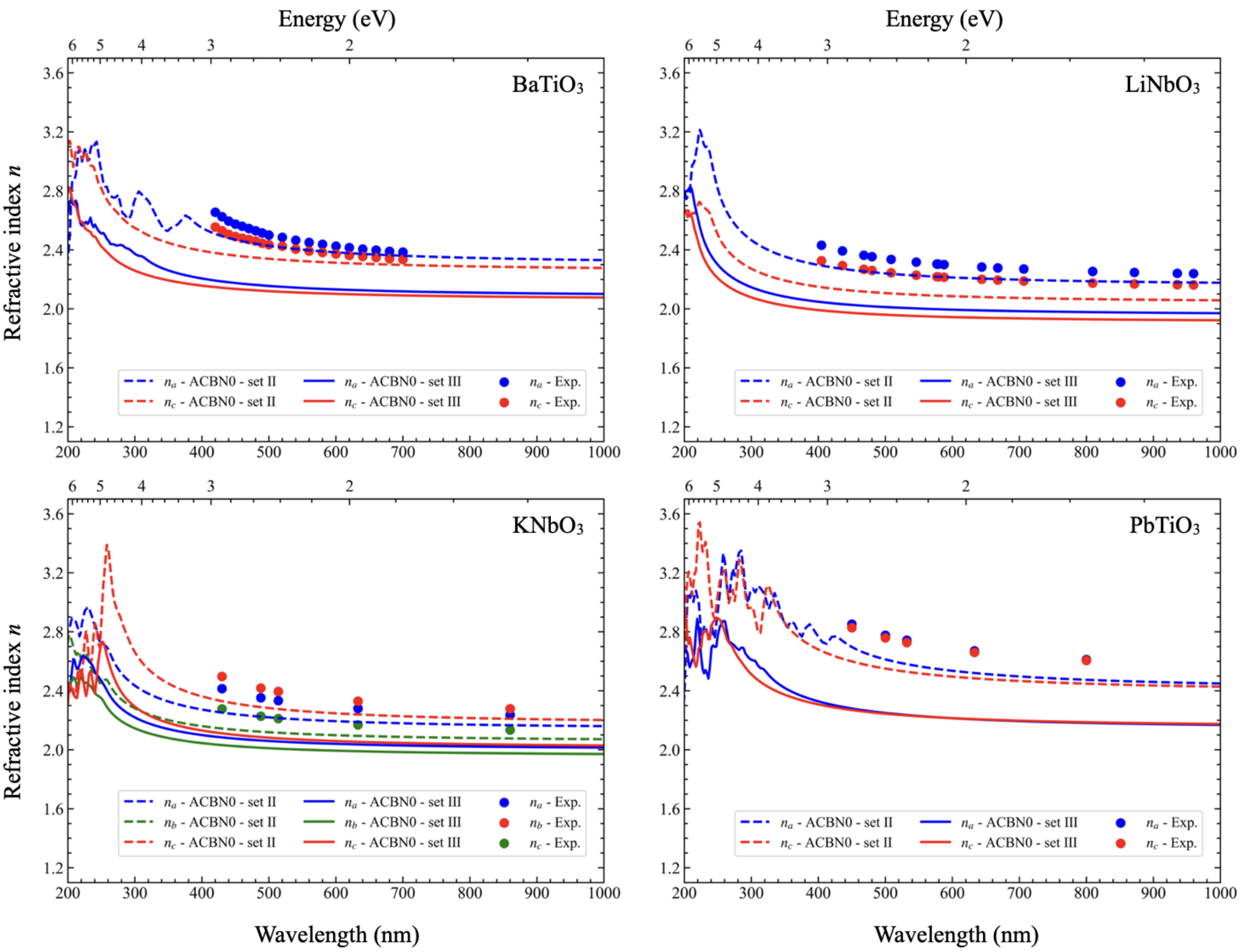

Figure S1: Refractive index dispersion in non-centrosymmetric oxides ($BaTiO_3$, $LiNbO_3$, $KNbO_3$, $PbTiO_3$) using ACBN0. Parameter sets II and III correspond to $\{U_{TM}, V_{TM-O}\}$ and $\{U_{TM}, U_O, V_{TM-O}\}$, respectively. Experimental values are as in [34–37].